# Ultrafast Spontaneous Motion of Nanodroplets

Cunjing Lv[1,2], Chao Chen[1,2], Yin-Chuan Chuang[3], Fan-Gang Tseng[3,4], Yajun Yin[1], Francois Grey[2], Quanshui Zheng[1,2*]

[1] Department of Engineering Mechanics, Tsinghua University, Beijing 100084, China

[2] Center for Nano and Micro Mechanics, Tsinghua University, Beijing 100084, China

[3] Department of Engineering and System Science, National Tsing Hua University, Hsinchu 30013, Taiwan

[4] Research Center for Applied Sciences, Academia Sinica, Taipei 11529, Taiwan

**Making liquid droplets move spontaneously on solid surfaces is a key challenge in lab-on-chip and heat exchanger technologies. The best-known mechanism, a wettability gradient, does not generally move droplets rapidly enough and cannot drive droplets smaller than a critical size. Here we report how a curvature gradient is particularly effective at accelerating small droplets, and works for both hydrophilic and hydrophobic surfaces. Experiments for water droplets on tapered surfaces with curvature radii in the sub-millimeter range show a maximum speed of 0.28 m/s, two orders of magnitude higher than obtained by wettability gradient. We show that the force exerted on a droplet scales as the surface curvature gradient. Using molecular dynamics simulations, we observe nanoscale droplets moving spontaneously at over 100 m/s on tapered surfaces.**

Most observations of the spontaneous motion of droplets on solid surfaces are variants of the Marangoni effect, due to wettability gradients (*1-4*). Many techniques, such as thermal (*4, 5*), chemical (*6*), electrochemical (*7*), and photochemical methods (*8, 9*), can be used to make a flat surface exhibit a continuously varying liquid contact angle. Droplets on such surfaces tend to move toward the region with lower contact angle. Although spontaneous motion has also been observed for droplets on substrates with uniform surface energy but varying surface roughness (*10*), its mechanism can still be

---

categorized as a Marangoni effect because of the varying effective contact angle. The speed of spontaneous droplet motion resulting from such effects is typically in the range of micrometers to millimeters per second (*5*), far too low for applications in areas such as liquid-based thermal management of fuel cells and semiconductor devices.

The main obstacle to droplet motion on a solid surface arises from contact angle hysteresis. In order to overcome this difficulty, large droplets are typically used and external sources of energy are introduced such as vibrating or heating the surface (*5, 11, 12*). Spontaneous motion up to 0.3 m/s was observed (*5*) for condensation droplets coming from saturated steam (100°C) on a silicon wafer with a radial gradient of surface energy. The current record for spontaneous droplet motion under ambient conditions is about 0.5 m/s (*13, 14*), using a chemically patterned and nanotextured surface.

Here we report a mechanism that can result in ultrafast spontaneous motion of micro- and nanoscale droplets. In contrast with the Marangoni effect, neither surface energy gradient nor surface roughness variation is required. The purely geometrical consequences of the surface curvature gradient provide the driving mechanism. First, we will present the experimental and simulation-based evidence for the effect. Then we will explain these results using a simple model, and show that it accounts quantitatively for both the increase of the velocity at the nanoscale, and the persistence of the effect on hydrophobic surfaces.

To observe the effect experimentally, tapered surfaces were sharpened from glass tubes of 1.5 mm in diameter by a flaming/brown micropipette puller (Model P-1000, Sutter Instrument). The video frames shown in Fig. 1A illustrate a typical spontaneous motion of a 1 mm$^3$ water drop under ambient conditions on the tapered surface. The droplet moves toward the direction of increasing cross-section radius, $R$, of the tapered tube. The velocity, $v$, of the droplet is measured as a function of $R$ along the tapered tube, plotted as red squares in Fig. 1B as the surface had been treated by oxygen plasma and then in MTS solution. For this surface, the water contact angle, $\theta$, and the contact angle hysteresis, $\Delta\theta$, defined as the difference between the advancing and receding contact angles of the droplet, are measured as about 0° and 1.5°, respectively. Other data in Fig. 1B show the velocity of identical droplets on the same tapered tube after cleaning by deionized water (green triangles, with $\theta \approx 41°$ and $\Delta\theta \approx 4°$) and then treated its surface in an oxygen plasma (blue circles, with $\theta \approx 5°$ and $\Delta\theta \approx 2°$). The velocity is not monotonic along the surface, but it can reach well over 0.2 m/s and it increases with decreasing contact angle. The maximum speed (~0.28 m/s) observed in these experiments is two orders of magnitude higher than those obtained by standard wettability gradients (*15*).

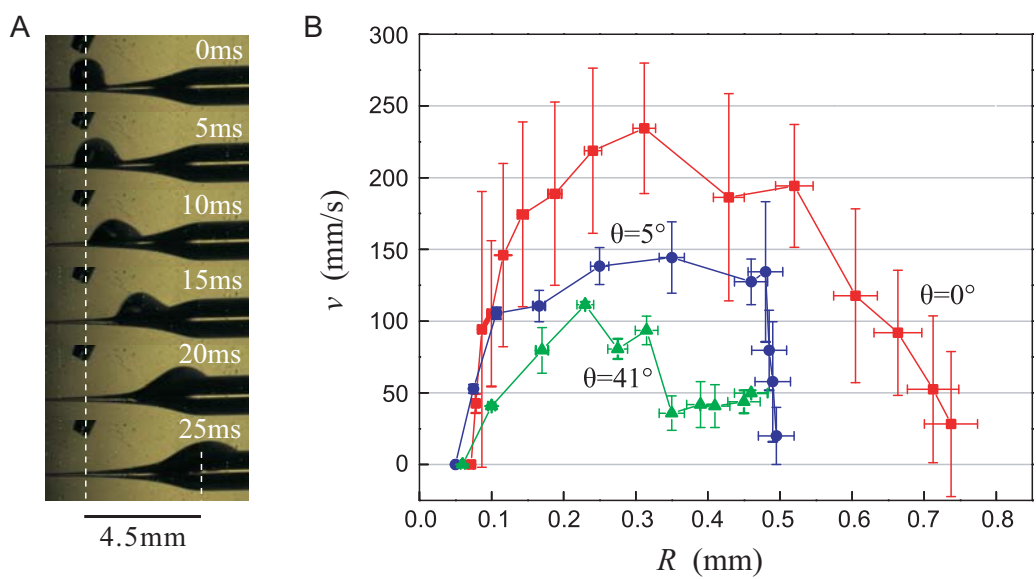

**Fig. 1.** Experimentally observed fast spontaneous motion of 1 mm³ water droplets on tapered glass rods with different surface conditions. (**A**) Video frames of a droplet moving on an oxygen plasma and MTS treated tapered glass surface with radius range 0.07 - 0.75 mm, contact angle $\theta \approx 0°$, and hysteresis $\Delta\theta \approx 1.5°$. (**B**) Velocity, $v$, as a function of local radius, $R$, of the glass tapered tube, for droplets moving on the same tapered surface with three different surface conditions: untreated ($\theta \approx 41°$, $\Delta\theta \approx 4°$); O₂ plasma-treated ($\theta \approx 5°$, $\Delta\theta \approx 2°$); and MTS nanotextured plus O₂ plasma-treated ($\theta \approx 0°$, $\Delta\theta \approx 1.5°$) (see the videos in the supplementary materials for more details).

We complemented these experiments with molecular dynamics (MD) simulations based on the LAMMPS platform (*16*), allowing us to study similar phenomena on the nanoscale. A nano-cone was modeled as a rigid framework of a monolayer of conically rolled graphene, as shown in Figs. 2A and 2C. The droplet was modeled by a standard code (SPC/E) (*17*) with the same parameters as given in Ref. 18. A Lennard-Jones potential, $\phi(r) = 4\varepsilon[(\sigma/r)^{12} - (\sigma/r)^6]$, was used to characterize the cone-liquid van der Waals interaction (*19*), where $r$ denotes the distance between atoms. By fixing the equilibrium distance $\sigma$ at 0.319 nm and adopting two different values, 5.85meV and 1.95meV, for the well depth $\varepsilon$, we obtained two different contact angles $\theta = 51°$ (hydrophilic) and 138° (hydrophobic). The model cone has a half-apex angle $\alpha = 19.5°$ and a height of 7 nm. As illustrated in Figs. 2A and 2C, we then cut off the tip at the heights of 1.5nm and 3.5nm, for modeling the motion of a 2nm-diameter droplet containing 339 molecules on the outer and inner cone surface, respectively. In the simulations, the temperature is kept at 300 K with a Nosé/Hoover thermostat, and the whole system is located in a finite vacuum box. Initially, the mass center of the droplet is fixed at a starting point for 100 ps, to reach thermal equilibrium. Then the droplet is released to move freely along the conical surface for the next 300 ps.

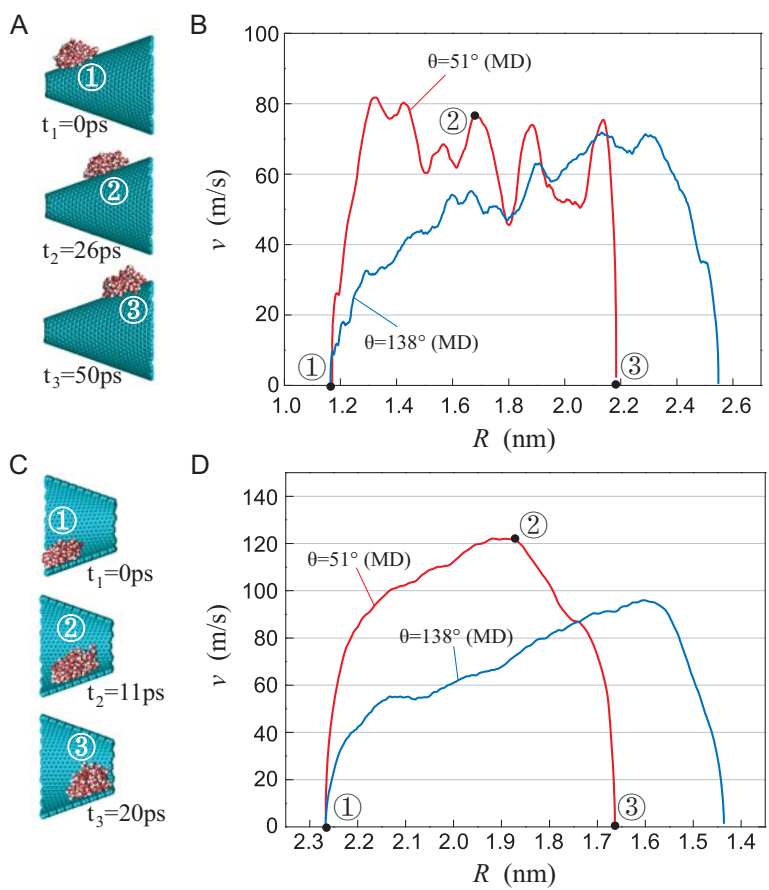

**Fig. 2.** Molecular-dynamics simulations of spontaneous motion of a droplet consisting of 339 water molecules on the outer surface (**A**), (**B**) and inner surface (**C**), (**D**) of a nano-cone. In (**B**) the velocity of the droplet, $v$, which starts near the narrow end of the cone, is plotted as a function of the cross-section radius of the cone, $R$, until it reaches the large end of the cone, where it briefly stops. The red (solid) and blue (dot-dashed) correspond to water contact angles of 51° and 138° respectively. In (**C**) the corresponding velocity of the droplet moving inside the cone is plotted, from a starting point near the large end. Illustrations in (**A**), (**C**) show droplet positions at three different times, $t_1$, $t_2$ and $t_3$, that are noted in the graphs (from the MD simulation movie, see supplementary materials).

When the droplet is released near the small end of the external surface, it starts to move spontaneously toward the larger end, as illustrated in Figs. 2A and 2B, with the maximum velocity reaching briefly over 80 m/s, until the droplet encounters the large end of the cone, where it bounces off the potential barrier. The red (solid) curve in Fig. 2B depicts the dependence of velocity, $v$, on radius of the cone $R$ measured directly below the center of mass of the droplet, for contact angle $\theta = 51°$. Motion is shown up until the point where the droplet instantaneously comes to rest at the potential barrier. Snapshots of the droplet at three times during this motion are illustrated in Fig. 2A. Similar spontaneous motion is observed for the hydrophobic case ($\theta = 138°$) a maximum velocity of 70 m/s being reached more slowly than in the hydrophilic case. When the droplet is released on the inner surface, we observe it to move spontaneously towards the smaller end of the cone, as illustrated in Figs. 2C and 2D. Even higher maximum velocities are reached in this case, of over 120 m/s for a hydrophilic interaction and over 90 m/s for the hydrophobic case.

To reveal the mechanism, we first explore how the surface energy of a droplet varies on a curved surface. For droplets that have diameters smaller than the capillary length, $\lambda_c = (\gamma/\rho g)^{1/2}$, where $\gamma$ and $\rho$ are the free surface tension and mass density of the liquid, and $g$ is the gravity acceleration, we can ignore the gravity potential. Thus, the total free energy can be quantified as (*1, 20*) $U = (A_{LV} - A_{LS}\cos\theta)$, where $A_{LV}$ and $A_{LS}$ denote the liquid-vapor and liquid-solid interface areas. For a droplet placed on a conical surface with half-apex angle, $\alpha$, we can use a finite element code, Surface Evolver (*21*), to deduce its shape from minimizing $U$ for a fixed droplet volume. The dots plotted in Fig. 3a show the resulting normalized surface free energy, $u = (\gamma A_s)^{-1}U$, versus $r_s\kappa$, where $r_s$ denotes the radius and $A_s = 4\pi r_s^2$ denotes the surface area of a spherical droplet of the same volume. $\kappa$ denotes the local curvature which is equal to $-R^{-1}\cos\alpha$ or $R^{-1}\cos\alpha$ when the droplet is placed on the external or internal conical surface, respectively, the local cross-section radius being *R*.

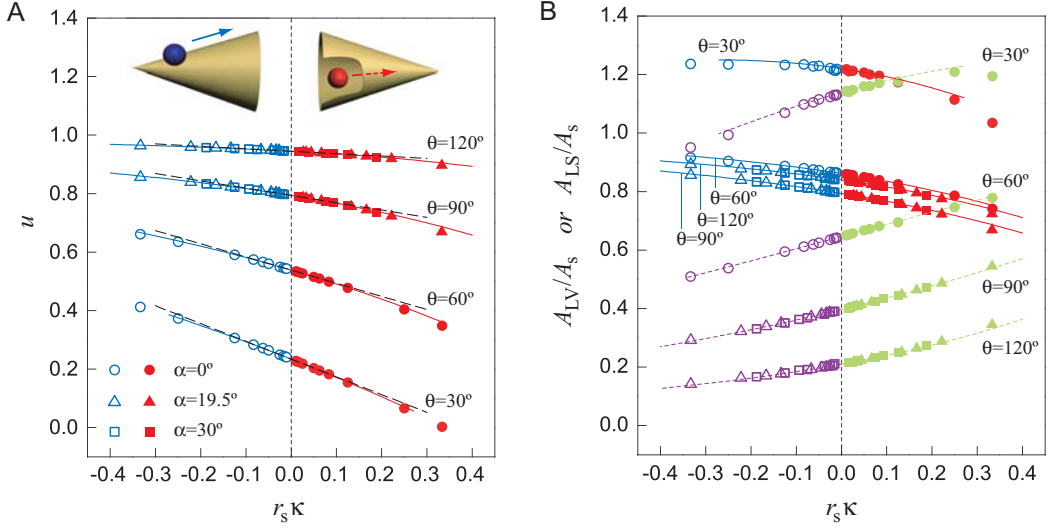

**Fig. 3.** Droplet free energy and interface area as a function of curvature. (**A**) Numerical results based on finite-element analysis for normalized total free energy $u$ of a small water droplet on a conical surface, as a function of local curvature, $\kappa$, measured directly below the center of mass of the droplet. Results for water contact angles $\theta = 120°$, $90°$, $60°$, and $30°$ are shown. For the droplet on the outer surface of the cone, this is a repulsive potential surface, relative to the cone apex, and for the droplet on the inside, it is attractive. The square and triangle symbols represent results for half-apex angles of the cone $\alpha = 30°$ and $19.5°$, respectively, and the circles represent results on cylinders ($\alpha = 0°$). The black dashed lines are plots of Eq. 2 with different contact angles. (**B**) The solid and dashed lines show how the liquid-vapor interface area $A_{LV}$ and the liquid-solid interface area $A_{LS}$ vary with the local curvature.

The results in Fig. 3A for various contact angles show that the surface free energy is always decreasing as the curvature increases, independent of whether the surface is hydrophilic or hydrophobic. Thus, the droplets always tend to move toward the larger or smaller end of the cone, whenever the droplets are on the external or internal surface of the cone, respectively. We call this driving force, which is purely the result of the curvature gradient, *curvi-propulsion*. A more detailed understanding of the origins of curvi-propulsion can be gained by considering how $A_{LV}$ and $A_{LS}$ vary with the curvature $\kappa$, as illustrated by the solid and dashed lines, respectively, in Fig. 3B. The dominant energetic effect is that $A_{LV}$ decreases monotonically with $\kappa$. Since $A_{LS}$ increases with $\kappa$, from Eq. 1 we see that this will either enhance or weaken the effect of decreasing $A_{LV}$ when the surface is hydrophilic ($\cos\theta > 0$) or hydrophobic ($\cos\theta < 0$), respectively. In both cases the net result, as shown in Fig. 3A, is that $U$ decreases monotonically with $\kappa$. Hence a droplet on the external cone surface moves towards $\kappa=0$ (away from the cone apex) and on the internal cone surface towards large positive $\kappa$ (towards the cone apex). This holds true for both hydrophilic and hydrophobic surfaces, though the net force, and hence the maximum velocity, will be smaller in the hydrophobic case.

The plotted results in Fig. 3A are scale-independent and are thus valid for different sized droplets and cones, from millimeter to nanometer. Furthermore, we find that the results are independent of half-apex angles, as numerically validated in Fig. 3A with $\alpha$ = 0° (circles for cylindrical tubes), $19.5^0$ (triangles), and $30^0$ (squares), and theoretically proved in the supplementary materials. Although an analytical relationship between $u$ and $r_s\kappa$ may not exist, its tangent line equation, $u = u_0 - \eta r_s\kappa$, at $r_s\kappa = 0$ can be shown to have the following coefficients:

$$u_0(\theta) = \sqrt[3]{(2+\cos\theta)\sin^4\frac{\theta}{2}}, \quad \eta(\theta) = \frac{1}{4}\cdot\frac{(1+\cos\theta)^2}{(2+\cos\theta)}, \tag{1}$$

where $u_0$ represents the normalized surface energy as a droplet is placed on a flat surface ($\kappa = 0$). Thus, a droplet on the cone surface is subject to a curvi-propulsion force, $F_c = -dU/ds$, that can be approximated as

$$F_c \approx \frac{3\gamma V}{2R^2}\eta(\theta)\sin 2\alpha, \tag{2}$$

where $V$ is the droplet volume and $s$ is a meridian coordinate pointing in the direction of increasing curvature.

Interestingly, the curvi-propulsion force $F_c$ scales as $R^{-2}$, analogous to an electrostatic force between two particles, with the effective potential between the droplet and the cone apex being repulsive on the outer surface, and attractive on the inner surface.

If hysteresis were negligible, then a droplet placed at any position, characterized by radius $R_0$, on a conical surface would always start spontaneous motion. From Eq. 2, we can obtain the explicit solution for the speed reached at position $R$ as follows:

$$v_{id}(R) = kv_c \left| \frac{\lambda_c}{R_0} - \frac{\lambda_c}{R} \right|^{1/2}, \qquad (3)$$

where $\lambda_c = (\gamma/\rho g)^{1/2}$ and $v_c = (\gamma g/\rho)^{1/4}$ are the capillary length and speed, respectively, and have the values $\lambda_c \approx 2.7$mm and $v_c \approx 163$mm/s for water at room temperature, while $k = [6\eta(\theta)\cos\alpha]^{1/2}$ is a monotonically decreasing function of $\theta$ with values $k = (2\cos\alpha)^{1/2}$ for $\theta = 0°$ and $k = 0$ for $\theta = 180°$. We call $v_{id}$ in Eq. 3 the *ideal* spontaneous motion speed, which is valid for droplets on both internal and external conical surfaces. On the external surface, the ideal speed has an upper limit $v_{up} = kv_c(\lambda_c/R_0)^{1/2}$ that is approachable as $R \to \infty$. Compared with the experimental and the numerical results shown in Figs. 1 and 2, the upper limits $v_{up}$ are equal to 0.8 m/s on the millimeter scale ($\theta \approx 41°$, $\alpha \approx 1°$, $r_s = 0.62$mm, $R_0 \approx 0.1$ mm) and 295 m/s on the nanoscale ($\theta \approx 51°$, $\alpha \approx 19.5°$, $r_s = 1$ nm, $R_0 = 1.2$nm). Since hysteresis and viscosity tend to reduce the ideal speed, we conclude that the above theoretical estimations agree well with our experimental observations and simulations.

Creeping motion of droplets on cones has been observed previously (*6, 22-25*) We distinguish the results described here from previous observations and analyses reported earlier, in several respects. Relatively large drops were considered in previous studies, which either encircled the cone externally or filled in a conical tube. The mechanism behind the creeping motion for such large drops is simply contact area gradient, while the driving mechanism for the spontaneous motion of the droplets we observe, which are small compared to the local dimensions of the cone, is curvature gradient. As a result, the wettability-independent motion always toward the larger or small end of the cone, for droplets on the external or internal conical surfaces, respectively, contrasts with larger drops either encircling or filling a cone, which may tend to move in different directions, depending upon the wettability (*22-25*). Above all, we show that droplets much smaller than considered previously can move spontaneously due to curvi-propulsion. At the nanoscale, we predict unprecedented speeds for the motion, confirmed by molecular dynamics simulations.

In conclusion, we have demonstrated ultra-high velocity motion of droplets moving on tapered surfaces, by experiment on the sub-millimeter scale and by molecular dynamics simulation on the nanoscale. The peak velocities achieved this way can exceed 100m/s on the nanoscale, orders of magnitude faster than any previously reported result for spontaneous droplet motion. We explain this remarkable effect, which we term curvi-propulsion, as being due to the much higher curvature gradients that nanoscale droplets can experience. We corroborate the effect using finite-element analysis and simple scaling arguments.

If suitable substrates can be designed to exploit this phenomenon, for example using arrays of nanoscale tapered structures, we speculate that curvi-propulsion could be useful for a range of practical applications where mass transport via droplet motion plays a key role, including rapid cooling, passive water collection and micro chemical synthesis. The curvi-propulsion mechanism may also prove useful in understanding biological phenomena such as inter-cell transport through cytonemes or filopedial bridges (*26*).

**Acknowledgements:** Financial support from the NSFC under grant No.10872114, No.10672089, and No. 10832005 is gratefully acknowledged. The authors greatly thank Professor David Quéré for a critical reading of the manuscript.

# Supplementary Materials for

## Ultrafast Spontaneous Motion of Nanodroplets

by Cunjing Lv, Chao Chen, Yin-Chuan Chuang, Fan-Gang Tseng, Yajun Yin, Francois Grey, Quanshui Zheng[1]

[1]To whom correspondence should be addressed. E-mail: zhengqs@tsinghua.edu.cn

**This PDF file includes:**

Materials and Methods
Supplementary Text
Figs. S1 to S 3
References

### S.1 Details of Experiments

**Materials:**

Glass capillaries with tip/end diameters of 140um/1.5mm were used. The tips of the capillaries were sharpened by a flaming/brown micropipette puller (Model P-1000, Sutter Instrument, U.S.A.) to obtain needle-shape capillaries.

Anhydrous alcohol (99.5%) and anhydrous toluene (99.9%) were purchased from Shimakyu's Pure Chemical (Osaka, Japan) and J. T. Baker (Phillipsburg, NJ, U.S.A). MTS (Methyltrichlorosilane, $CH_3SiCl_3$, 99% purity, catalogue number: M85301) was purchased from Sigma-Aldrich, USA.

**Methods:**
1. *Surface cleaning*

   Acetone and IPA were used to pre-clean surfaces of the glass capillaries. Then deionized (DI) water with a resistance of 20MΩ was used to carefully clean the surfaces.

2. *Surface Modification process*

   Outer surfaces of all the capillaries were pretreated by oxygen plasma at 100 W and 75 mtorr chamber pressure for 300 s. Then the capillaries were put in 0.014 M MTS solution in anhydrous toluene for 75 min under 23°C and 75 % RH environmental conditions. For removing the residual molecules that were not immobilized from the surface, the capillaries were then rinsed in anhydrous toluene, ethanol, a mixed solution of ethanol and DI (1:1), and DI water in a sequence. Compressed dry air was employed to blow dry the capillaries, and finally the capillaries were put into an oven for annealing at 120°C for 10 min [1].

### 3. Measurement

The static contact angle (CA) and contact angle hysteresis (CAH) of each surface before and after treatment were measured with FTA 200 instrument (First Ten Angstroms, U.S.A.) by using pipette to drop 2 $\mu$L DI water on a flat glass slide (Kimble glass, Owen-Illinois, U.S.A.) instead of the glass capillaries because of the difficulties in defining contact angle measurement on a curved surface.

Images of the moving 1 $\mu$L droplet were captured with a high speed camera (Fastcam-Ultima APX, Photron) coupled with a long focus optical system (Zoom 700X with internal focus vertical illuminator, OPTEM International U.S.A).

**Results & Discussions:**

After basic cleaning, the plain glass capillary surface showed a CA of 41° and CAH of 4°. Then it became highly hydrophilic with a CA of 5° and CAH of 2° after oxygen plasma treatment.

The capillary surface was modified with 3D nanotexture made of MTS aggregations (*1, 2*) and treated by oxygen plasma. It finally presented a CA of 0° and CAH of 1.5°. (as shown in Fig. S1)

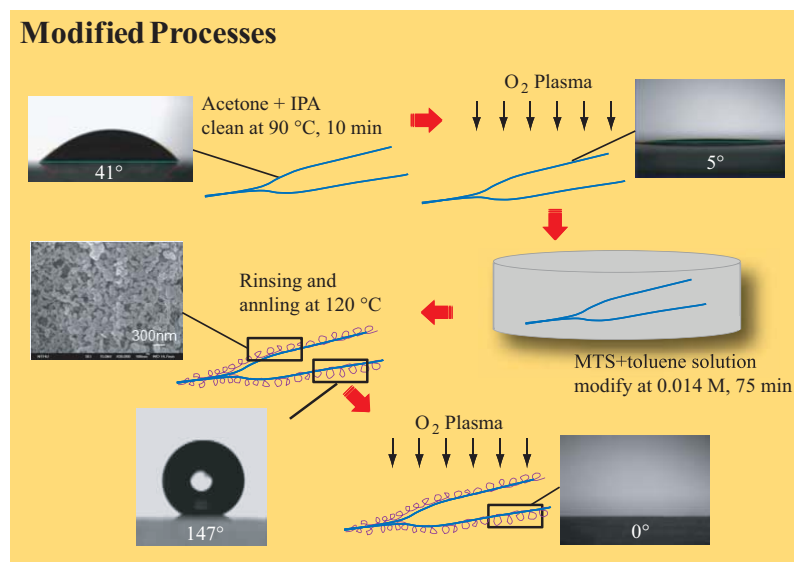

**Fig. S1.** Chemical modification processes of the capillary surface.

## S.2 Theory

### S.2.1 Relationship between surface energy and curvature for a water droplet

The liquid-vapor interface energy, $\gamma$, specifies two parameters, the capillary length $\lambda_c = (\gamma/\rho g)^{1/2}$ and capillary speed $v_c = (\gamma g/\rho)^{1/2}$, where $\rho$ is the liquid mass density and $g$ is the gravity constant. For water at the room temperature, we have $\lambda_c \approx 2.7$mm and $v_c \approx 163$mm/s.

For a droplet placed on a solid surface, if the droplet size is smaller than the capillary length, we can ignore the influence of gravity and express the total free

energy as:

$$U = A_{LV}\gamma + A_{LS}(\gamma_{LS} - \gamma_{SV}) = \gamma(A_{LV} - A_{LS}\cos\theta) \quad (S1)$$

where $A_{LV}$ and $A_{LS}$ denote the liquid-vapor and liquid-solid interface areas, $\gamma_{LS}$ and $\gamma_{SV}$ the liquid-solid and solid-vapor interface energies, and $\theta$ is the Young's contact angle which is associated with the interface energies in the form $\cos\theta = (\gamma_{SV} - \gamma_{LS})/\gamma$.

For a given volume, a droplet adopts a shape for which the total free energy $U$ is a minimum. For a droplet placed on a flat sphere surface, the droplet's shape is a spherical cap. For a droplet placed on a conical surface, there is no simple analytical solution for the shape. Hereinafter we show that we can find an analytical approximation.

From the theory of differential geometry (3), it is known that locally, the departure of a curved surface from it tangent plane is determined by the two principal curvatures of the surface, $\kappa_1$ and $\kappa_2$. Therefore, if the droplet's size is smaller than the curvature radii $1/|\kappa_1|$, $1/|\kappa_2|$, then the free energy $U$ depends only upon the local property of the surface, namely $\kappa_1$ and $\kappa_2$: $U = U(\kappa_1,\kappa_2)$. Mathematically, denoting by $r_s$ the radius of a spherical droplet with the same volume, we can write the small droplet condition as $r_s\kappa \ll 1$, where $\kappa = \max\{|\kappa_1|,|\kappa_2|\}$. Thus, we can take the Taylor expansion of $U(r_s\kappa_1, r_s\kappa_2)$ as follows:

$$U(r_s\kappa_1, r_s\kappa_2) = U(0,0) + \frac{\partial U(0,0)}{\partial(r_s\kappa_1)}r_s\kappa_1 + \frac{\partial U(0,0)}{\partial(r_s\kappa_2)}r_s\kappa_2 + O((r_s\kappa)^2). \quad (S2)$$

Since water is isotropic, this yields

$$\frac{\partial U(0,0)}{\partial(r_s\kappa_1)} = \frac{\partial U(0,0)}{\partial(r_s\kappa_2)}. \quad (S3)$$

Consequently,

$$U(r_s\kappa_1, r_s\kappa_2) = U(0,0) + \frac{\partial U(0,0)}{\partial(r_s\kappa_1)} 2r_s H + O((r_s\kappa)^2), \quad (S4)$$

where $H = (\kappa_1 + \kappa_2)/2$ is the mean curvature. In other words, if we ignore higher order terms, then $U$ depends upon only the mean curvature, rather than the Gauss curvature $K = \kappa_1\kappa_2$.

Based on the above result, we can find an analytical solution of $U(R)$ for droplets placed on a spherical surface with radius $R$, we then replace $R^{-1}$ by the mean curvature $H$ for other curved surfaces. In particular, for a cone with the half apex angle $\alpha$ and local cross-section radius $R$, the two principal curvatures are $\kappa_1 = R^{-1}\cos\alpha$ and $\kappa_2 = 0$, respectively. We can replace $R^{-1}$ in $U(R)$ for the sphere with $H = (R^{-1}\cos\alpha)/2$ for the cone.

**S.2.2 Surface energy of water droplet on a spherical surface**

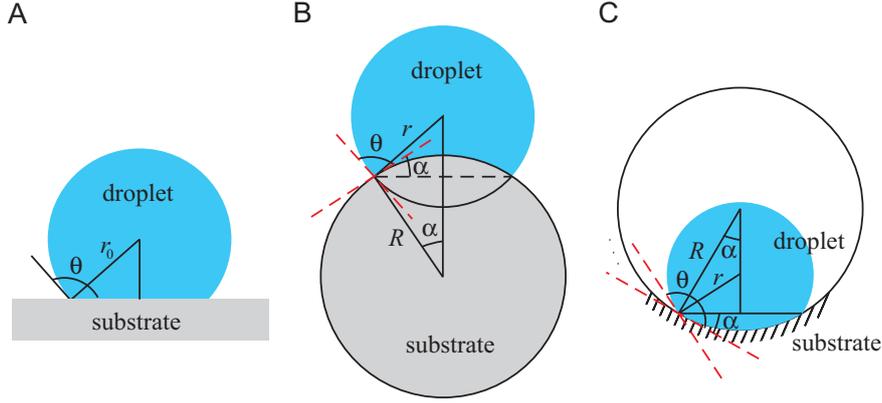

**Fig. S2.** Sketch of the wetting states of the same water droplet on a flat (**A**), outer (**B**) and inner (**C**) part of a curved surface, respectively.

For a droplet placed on a flat surface (Fig. S2A), the free energy can be easily solved as:

$$U_0 = \pi r_0^2 (2 - 3\cos\theta + \cos^3\theta)\gamma = \frac{3V\gamma}{r_0}, \tag{S5}$$

where

$$\begin{aligned} V &= \frac{\pi}{3} r_0^3 \left(2 - 3\cos\theta + \cos^3\theta\right) \\ &= \frac{\pi}{3} r_0^3 (1-\cos\theta)^2 (2+\cos\theta), \\ &= \frac{4\pi}{3} r_0^3 (2+\cos\theta) \sin^2\frac{\theta}{2} \end{aligned} \tag{S6}$$

is the droplet volume, and $r_0$ is the radius of the water-solid contact area. Denoting by $r_s$ the radius of a spherical droplet of the same volume $V$, from Eq. (S6) we can further deduce

$$\frac{r_s}{r_0} = \sqrt[3]{\frac{2-3\cos\theta+\cos^3\theta}{4}} = \sqrt[3]{(2+\cos\theta)\sin^4\frac{\theta}{2}}. \tag{S7}$$

When a water droplet with the same volume is placed on the external spherical surface (see Fig. S2B), referred to Eq. (S6) and Fig. S2B we can get the following relationships:

$$\begin{cases} V = \frac{\pi}{3} r^3 (1-\cos\hat{\theta})^2 (2+\cos\hat{\theta}) - \frac{\pi}{3} R^3 (1-\cos\alpha)^2 (2+\cos\alpha), \\ R\sin\alpha = r\sin\hat{\theta}, \end{cases} \tag{S8}$$

where $r$ is the radius of the water droplet on the spherical surface, $R$ is the radius of the contact area, $\hat{\theta} = \theta + \alpha$, and $\alpha$ as shown in Fig. S2B is the half-spanned angle of the wetted area of the sphere. The two interface areas are:

$$A_{LV} = 2\pi r^2 (1-\cos\hat{\theta}), \quad A_{LS} = 2\pi R^2 (1-\cos\alpha), \tag{S9}$$

From Eqs. (S1), (S8), and (S9) we can numerically solve $U$ as a function of the mean curvature, plotted as the smooth solid and dotted lines in Fig. 3AB. The excellent agreement with the numerical results for cone surfaces proves that ignoring the dependence of $U$ upon the Gauss curvature is a very good approximation. Similar results can be obtained as the water droplet is placed on the internal surface of the sphere (see Fig. S2C).

### S.2.3 Linear relationship between mean curvature $\kappa$ and surface energy $U$

There is no simple analytical solution for the shape of a droplet on an arbitrary curved substrate (*4, 5*), So, the way we deduce the relationship between the surface free energy $U$ and the substrate curvature $\kappa$ is to use a first-order linear expansion of the surface free energy on a spherical surface. Based on Eq. 1 and Fig. S2B, we obtain

$$\frac{dU}{d\kappa} = \frac{d}{dR}\left[\gamma\left(A_{LV} - A_{LS}\cos\theta\right)\right] \cdot \frac{dR}{d\kappa}$$
$$= 4\pi\gamma R^2 \left\{ r\left(1-\cos\hat{\theta}\right)\frac{dr}{dR} + \frac{1}{2}\left(r^2\sin\hat{\theta} - R^2\sin\alpha\cos\theta\right)\frac{d\alpha}{dR} - R\left(1-\cos\alpha\right)\cos\theta \right\}$$

(S10)

In Eq. (S10), since for a droplet placed on the external spherical surface, the mean curvature is $\kappa = -1/R$, we have $dR/d\kappa = R^2$. Based on the conservation relationship Eq. (S8), after some calculation, we obtain:

$$\begin{cases} \dfrac{dr}{dR} = \left(\dfrac{R}{r}\right)^2 \cdot \dfrac{\left(2-3\cos\alpha+\cos^3\alpha\right)}{\left(2-3\cos\hat{\theta}+\cos^3\hat{\theta}\right)} \\ \dfrac{d\alpha}{dR} = \dfrac{\sin\alpha - \dfrac{dr}{dR}\cdot\sin\hat{\theta}}{r\cos\hat{\theta}-R\cos\alpha} \end{cases}, \quad (S11)$$

As $\alpha \ll 1$, by developing a Taylor series

$$\cos\alpha\big|_{\alpha\to 0} = 1 - \frac{\alpha^2}{2} + \frac{\alpha^4}{24} + \cdots, \quad (S12)$$

$$\sin\alpha\big|_{\alpha\to 0} = \alpha - \frac{\alpha^3}{6} + \cdots, \quad (S13)$$

and ignoring higher order terms in $\alpha$, it is straightforward to obtain:

$$\frac{dU}{d\kappa} = -\frac{1}{2}U_0 r_0 \frac{(1+\cos\theta)^2}{(2+\cos\theta)}, \quad (S14)$$

Introducing the normalized free energies

$$u(\theta) = \frac{U}{\gamma A_s}, \quad u_0(\theta) = \frac{U_0}{\gamma A_s}, \quad (S15)$$

where $A_s = 4\pi r_s^2$ denotes the surface energy of a spherical droplet of the same volume, we have from Eqs. (S5) and (S7)

$$u_0(\theta) = \sqrt[3]{(2+\cos\theta)\sin^4\frac{\theta}{2}}, \tag{S16}$$

From Eqs. (S1), (S10)-(S14), we obtain the tangent relation of the normalized surface free energy as follows:

$$u = u_0 + 2\eta\left(\frac{r_s}{R}\right) = u_0 - 2\eta r_s \kappa, \tag{S17}$$

where

$$\eta(\theta) = \frac{1}{4}\cdot\frac{(1+\cos\theta)^2}{(2+\cos\theta)}. \tag{S18}$$

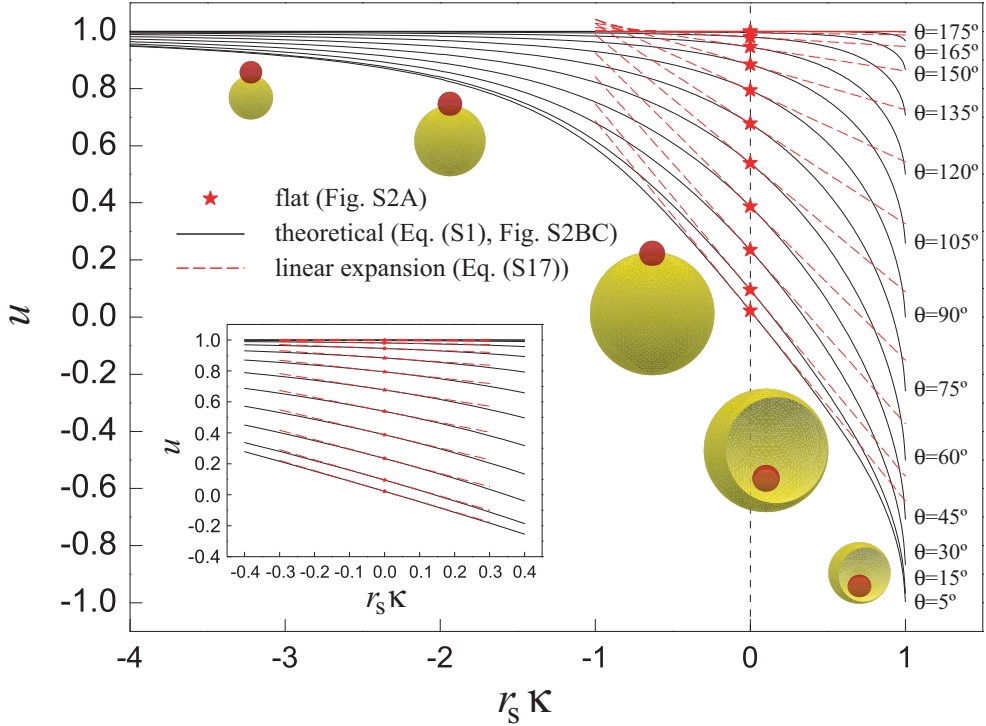

**Fig. S3.** The surface free energy $U$ of a droplet outside and inside of a spherical surface, a comparison between Eq. (S1) (black solid lines) and Eq. (S17) (red dashed lines) with different contact angles. The red five-pointed stars represent a droplet on flat surface (see Fig. S2A). Normalized curvature $r_s\kappa \in [-0.4, 0.4]$ in the insert.

From Fig. S3, we can see that our theoretical linear expansion Eq. (S17) is consistent with the theoretical results for a droplet on a spherical surface with small normalized curvature $r_s\kappa$. For a cone we should replace $\kappa = -1/R$ in Eq. (S17) by the mean curvature $\kappa/2 = -1/R$. The above analysis finally results in:

$$u = u_0 - \eta r_s \kappa. \tag{S19}$$

## S.3 Supplementary Movies

**Movie 1 (S1.mov, 1.74MB) –Side view of a 1$\mu$L water droplet move spontaneously on a coincal surface treated with MTS nanotexture plus O$_2$ plasma.**

When a 1.0 $\mu$L water droplet is released gently enough from the syringe, it will move spontaneously on the conical surface (see Fig. 1A). Glass capillaries with tip/end diameters of 140um/1.5mm and surface condition is MTS nanotextured plus O$_2$ plasma treatment with $\theta \approx 0°$, $\Delta\theta \approx 1.5°$. The actual speed of the water droplets in the experiment is 200 times higher than shown in the videos.

**Movie 2 (S2.mov, 273KB) –Side view of a 1$\mu$L water droplet moving spontaneously on a conical surface treated with O$_2$ plasma.**

When a 1.0 $\mu$L water droplet is released gently from the syringe, it will move spontaneously on the conical surface. The surface is O$_2$ plasma treated with $\theta \approx 5°$, $\Delta\theta \approx 2°$. The actual speed of the water droplets in the experiment is 10 times that shown in the video.

**Movie 3 (S3.mov, 498KB) – Side view of a 1$\mu$L water droplet moving spontaneously on a plain coincal surface.**

When a 1.0 $\mu$L water droplet is released gently from the syringe, it will move spontaneously on the conical surface. The surface condition is plain with $\theta \approx 41°$, $\Delta\theta \approx 4°$. The actual speed of the water droplets in the experiment is 10 times that shown in the videos.

**Movie 4 (S4.mov, 3.36MB) – Molecular dynamics simulation results for water droplet spontaneous motion on graphene cone**

The water droplet on the outer/inner conical surface is spontaneously moving toward the larger/smaller cross-section area regardless of the value of the water contact angle ($\theta$=50.7° (hydrophilic), 138° (hydrophobic), or nearly 180° (superhydrophobic)). The droplet bounces back when it meets the cone edge.

(2001).